\documentclass[11pt]{article}

\usepackage[utf8]{inputenc}
\usepackage[T1]{fontenc}
\usepackage{lmodern}
\usepackage[margin=1in]{geometry}
\usepackage{amsmath,amssymb}
\usepackage{gensymb}
\usepackage{graphicx}
\usepackage{xcolor}
\usepackage[font=small,labelfont=bf,labelsep=period]{caption}
\usepackage{authblk}
\usepackage[numbers,sort&compress]{natbib}
\usepackage[colorlinks=true,linkcolor=blue,citecolor=blue,urlcolor=blue]{hyperref}

\title{Thin-film lithium niobate integrated photonics at 493\,nm}

\author[1,$\dagger$,*]{Nicholas Achuthan}
\author[1,$\dagger$]{Let\'icia Magalh\~{a}es}
\author[1,$\dagger$]{Xudong Li}
\author[2]{Hugo Voncken}
\author[1]{Hana Warner}
\author[1]{Sophie Ding}
\author[1]{Andrea Cordaro}
\author[1]{Xinrui Zhu}
\author[3]{Keith Powell}
\author[1]{Anna Shelton}
\author[1]{Dylan Renaud}
\author[1]{Daniel Assumpcao}
\author[1]{Xiangying Zuo}
\author[4]{Michael Pushkarsky}
\author[4,**]{Amirhassan Shams-Ansari}
\author[1,***]{Marko Lon\v{c}ar}

\affil[1]{John A. Paulson School of Engineering and Applied Sciences, Harvard University, Cambridge, MA 02138, USA}
\affil[2]{Faculty of Applied Sciences, Delft University of Technology, Lorentzweg 1, 2628 CJ Delft, The Netherlands}
\affil[3]{Lumina Corp, Cambridge, MA 02138, USA}
\affil[4]{Monarch Quantum, San Diego, CA 92131, USA}
\affil[$\dagger$]{These authors contributed equally to this work.}
\affil[*]{\href{mailto:nachuthan@g.harvard.edu}{nachuthan@g.harvard.edu}}
\affil[**]{\href{mailto:amirhassan.shams-ansari@monarchquantum.com}{amirhassan.shams-ansari@monarchquantum.com}}
\affil[***]{\href{mailto:loncar@seas.harvard.edu}{loncar@seas.harvard.edu}}

\date{}

\begin{document}

\maketitle

\begin{abstract}
Visible and near-ultraviolet light underpins applications from biophotonics to quantum computing, which demand fast control of optical amplitude and phase. Integrated photonics offers scalable control, yet most visible-wavelength platforms lack gigahertz-scale modulation. We demonstrate a thin-film lithium niobate platform at 493~nm integrating grating couplers, beamsplitters, ring resonators, and modulators. Segmented traveling-wave modulators achieve a half-wave voltage--length product of 450~mV$\cdot$cm, $>$20~dB extinction, and bandwidth beyond 50~GHz. Ring resonators show $2.9 \pm 0.2$~dB/cm propagation loss and quantify the platform's photorefractive response. This enables agile visible beam delivery for trapped-ion systems, although photorefraction and substrate uniformity remain key challenges to scaling.
\end{abstract}

\section{Introduction}
\label{sec:introduction}

Visible and near-ultraviolet light enables a growing range of technologies, including quantum control of trapped ions and neutral atoms, optical atomic clocks, optogenetic neural stimulation, augmented-reality displays, and underwater optical communication \cite{park2024technologies, ludlow2015optical, mohanty2020reconfigurable, notaros2023liquid}. Most of these applications rely on precise modulation of visible light, which today is provided largely by bulk acousto-optic and electro-optic modulators and spatial light modulators that are large, power-hungry, and difficult to scale to many channels \cite{park2024technologies}. 

Optically controlled atomic systems, in particular, rely critically on laser fields for cooling, trapping, state preparation, coherent manipulation, and readout, and on modulators that gate, shape, frequency-shift, and stabilize these fields \cite{park2024technologies, montifiore2026blue}. However, as atomic systems increase in scale and complexity, generating and distributing these control fields using bulk optical components becomes increasingly difficult. Photonic integrated circuits (PICs) provide a route to miniaturizing this optical infrastructure by combining beam routing, splitting, frequency synthesis, and modulation within a compact and scalable architecture.  Trapped ions also impose some of the most demanding requirements on photonic integration: the optical control stack must combine visible and near-UV operation with gigahertz-scale agility, required for resolving hyperfine structure and driving Raman transitions \cite{ransford2025helios, dietrich2009barium, bruzewicz2019trapped}. 

\begin{figure}[htbp]
\centering\includegraphics[width=5.25in]{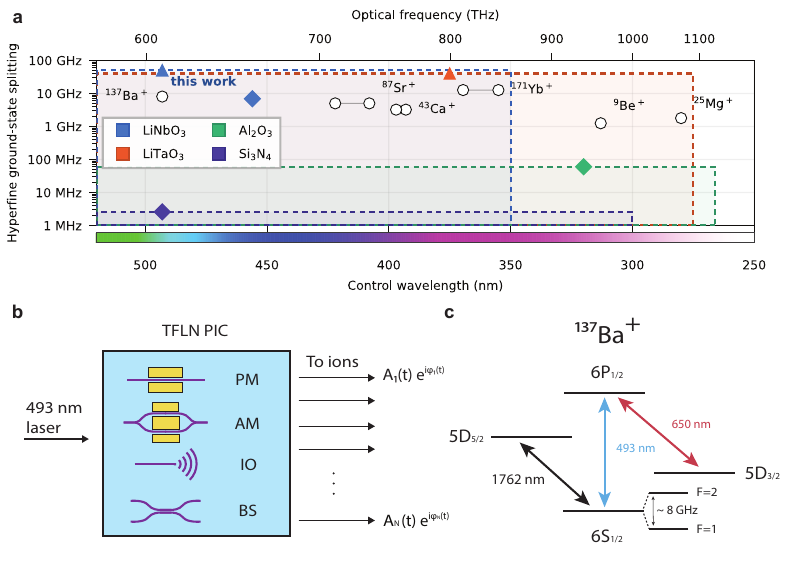}
\caption{\textbf{TFLN for controlling ions.} (a) Integrated photonic modulators operating below 500 nm, plotted by operating wavelength and demonstrated modulation bandwidth: TFLN at 493 nm (this work) and 456 nm \cite{celik2024roles}, silicon nitride at 493 nm \cite{montifiore2026blue}, thin-film lithium tantalate (TFLT) at 375 nm \cite{lin2026thin}, and alumina at 320 nm \cite{castillo2026cmos}. Triangles denote instrument-limited measurements, where higher bandwidth is expected, while diamonds denote measured roll-offs. White circles give the hyperfine ground-state splitting of leading ion species, plotted against the wavelengths their optical control requires. Each shaded region extends up to the highest modulation frequency demonstrated on that platform in this wavelength range, and rightward to the material's transparency edge. (b) Building blocks of the 493 nm electro-optic platform on TFLN. Phase modulation (PM), amplitude modulation (AM), input/output (IO), and beamsplitter (BS) functionality enables a single laser source to drive $N$ channels with programmable amplitude ($A_i$) and phase ($\phi_i$). (c) Basic level structure of $^{137}\mathrm{Ba}^+$. }
\label{fig1}
\end{figure}

Trapped ions stand to benefit greatly from photonic integration \cite{niffenegger2020integrated, mehta2020integrated, zhao2025integrated}, and, to date, trap-integrated photonics has enabled polarization-gradient cooling \cite{corsetti2026integrated, clements2026sub} and on-chip fluorescence collection \cite{knollmann2025collection}. Integrated modulators have also driven high-fidelity trapped-ion qubit operations, albeit only at megahertz-scale bandwidths \cite{hogle2023high, menssen2023scalable, montifiore2026blue}. While a large number of visible-wavelength photonic platforms deliver broad optical transparency with low loss \cite{west2019low, corato2024absorption}, integrated modulators operating below 500 nm show a clear trade-off between optical transparency and modulation speed [Fig.~\ref{fig1}(a)]. Low-loss platforms such as silicon nitride and alumina transmit well into the ultraviolet \cite{corato2024absorption, castillo2026cmos}. However, they lack an intrinsic electro-optic effect and rely on stress-optic or piezo-optomechanical modulation, which has so far remained below 100 MHz \cite{castillo2026cmos, montifiore2026blue}. Liquid-crystal-based modulators, meanwhile, are more compact still but switch at sub-kilohertz rates \cite{notaros2022integrated}. These speeds are more than an order of magnitude below the gigahertz-scale hyperfine splittings of the ion species of interest, and fast modulation at short wavelengths has been identified as a key missing capability for quantum applications \cite{park2024technologies}. 

Electro-optic thin films close this gap. Thin-film lithium niobate (TFLN) combines a broad transparency window with a strong Pockels effect \cite{zhu2021integrated}, making it a promising platform for integrated control of trapped ions. Low-loss visible TFLN waveguides \cite{desiatov2019ultra} and gigahertz-bandwidth modulators in the red and near-infrared \cite{renaud2023sub, celik2022high, assumpcao2024thin} have resulted in demonstrations of modulators spanning the visible band, extending to around 450 nm \cite{xue2023full, liu2024foundry, celik2024roles}. While the transparency window of TFLN is narrower than that of some alternatives, it nonetheless covers key control frequencies of several leading species, including  $\mathrm{Ba}^{+}$, $\mathrm{Sr}^{+}$, $\mathrm{Ca}^{+}$, and $\mathrm{Yb}^{+}$ [Fig.~\ref{fig1}(a)]. Furthermore, thin-film lithium tantalate (TFLT), an electro-optic film with a wider transparency window \cite{juvalta2006electro}, may extend this capability further into the ultraviolet, with a recent 375 nm modulator demonstration reaching comparable GHz-scale bandwidths \cite{lin2026thin}. 

Shorter wavelengths offer intrinsic advantages for electro-optics: the phase shift induced by the Pockels effect scales inversely with wavelength, and tighter optical confinement permits smaller electrode gaps without excess metal-induced loss \cite{renaud2023sub}. Visible wavelengths also pose unique challenges: feature sizes approach fabrication limits, scattering losses increase dramatically \cite{payne1994theoretical, corato2024absorption}, the higher optical group index complicates velocity matching with the microwave drive, and parasitic nonlinearities such as photorefraction can be enhanced \cite{suntsov2026photorefractive, yeh2025interface, holzgrafe2024relaxation}, producing drifts, the mechanisms of which are not fully understood \cite{park2024technologies}. DC bias drift is a further concern for blue-light modulators \cite{celik2024roles}.

In this paper, we develop a library of passive and active TFLN components operating at 493 nm, fabricated at the wafer scale on X-cut thin films. We envision an architecture where a laser is coupled to a TFLN PIC containing phase modulators, amplitude modulators, input/output couplers, and beam splitters [Fig.~\ref{fig1}(b)]. Together, these components distribute the input among multiple output channels while providing independent control over the amplitude $A_i(t)$ and phase $\phi_i(t)$ of each optical field. Such control enables a single laser source to address multiple ions with independently programmable optical waveforms. We specifically target quantum control of $^{137}\mathrm{Ba}^{+}$ ions [Fig.~\ref{fig1}(c)], whose ground-state hyperfine qubit, split by about 8 GHz, is used in current commercial trapped-ion processors, with the 493 nm $6S_{1/2}$--$6P_{1/2}$ transition used for state detection \cite{ransford2025helios}.

We first establish the passive platform, consisting of single-mode waveguides with a propagation loss of $2.9 \pm 0.2$ dB/cm, low-loss multimode-interference (MMI) beamsplitters, grating couplers, and directional couplers with integrated thermo-optic tuning to correct fabrication-induced splitting errors. This loss lies within the same range as that of high-confinement silicon nitride (1.2 dB/cm at 488 nm and 2.4 dB/cm at 461 nm) \cite{corato2024absorption} and ALD alumina (1.4--1.8 dB/cm at 405 nm) \cite{west2019low}. Lower losses, such as 0.24 dB/cm at 493 nm, have been achieved in low-confinement silicon nitride, but at the cost of bend radii of hundreds of micrometers and, without a Pockels effect, megahertz-range modulation bandwidths \cite{montifiore2026blue}. We then demonstrate traveling-wave phase modulators whose segmented electrodes halve the frequency-dependent RF loss and enable velocity matching, yielding an electro-optic bandwidth beyond 50 GHz. Using the same electrode design, push-pull amplitude modulators reach a half-wave voltage-length product of 450 mV$\cdot$cm and an extinction ratio of 20.4 dB, and generate carrier-suppressed bichromatic fields suited to Raman and motional-sideband control of trapped ions. Finally, we use ring resonators to quantify the photorefractive response of the platform. Although we do not yet operate these devices with trapped ions, they form the building blocks of the multichannel control architecture in Fig.~\ref{fig1}(b), and our results identify the challenges that remain before such integration.

\section{Device fabrication}
\label{sec:fabrication}

Our devices are fabricated on 6 inch LNOI X-cut wafers with 200 nm LN film and 7 $\mathrm{\mu m}$ of bottom oxide. We start by patterning optical layers using 150 keV electron beam lithography (ELIONIX ELS-BODEN 150) on MaN series EBL resist (when optimizing for loss as in ring resonator measurements, HSQ resist is used). The developed resist is transferred into the LN film with a 125 nm deep physical $\mathrm{Ar^+}$ etch, leaving a ridge waveguide on a 75 nm slab. Redeposition from the etch is removed with a heated SC-1 clean, after which the wafer is annealed in $520^\circ$C oxygen atmosphere for 2 hours to remove electron beam induced defects. The devices are cladded with 1 $\mathrm{\mu m}$ of $\mathrm{SiO_2}$ via chemical vapor deposition (Oxford Instruments PlasmaPro 100 ICPCVD) and annealed again \cite{shams2022reduced}. For active devices, electrodes are patterned with 50 keV electron beam lithography. The cladding is etched through in the exposed area so that electrodes can be deposited directly on the LN slab to improve microwave-optical field overlap. Electron beam evaporation is used to deposit 800 nm of Au on a 15 nm Ti adhesion layer, followed by a standard liftoff process. A heater layer is patterned via optical lithography (Heidelberg Instruments MLA150) and deposited via electron beam evaporation of 100 nm of NiCr. An additional layer of gold is defined to route the heaters to electrically-accessible contact pads. The wafer is then annealed at $200^\circ$C on a hot plate overnight, after which devices are either tested directly or singulated via a deep-etch process \cite{renaud2024mitigating}.

\section{Passive components}
\label{sec:passive}

\begin{figure}[htbp]
\centering\includegraphics[width=5.25in]{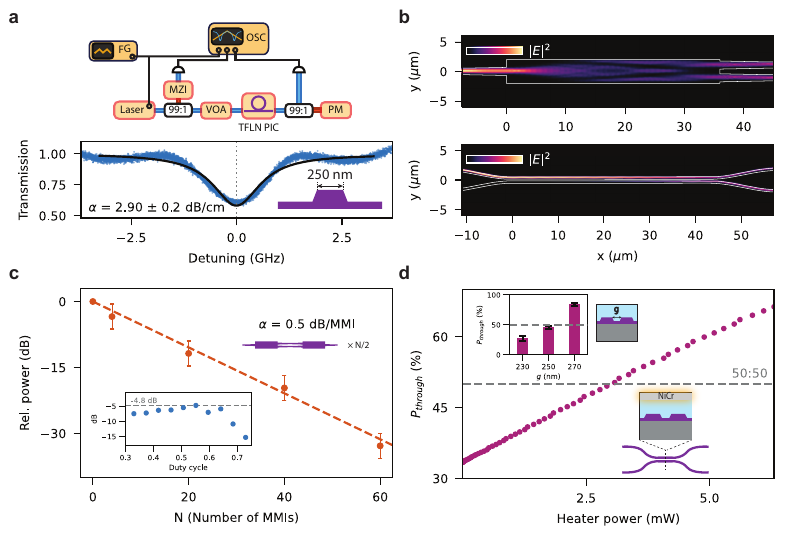}
\caption{\textbf{Passive components at 493 nm.} (a) Top: experimental setup used to characterize TFLN ring resonators at 493 nm. The laser frequency is scanned with a function generator while synchronous calibration MZI fringes and ring transmission signals are recorded. Bottom: normalized transmission of an undercoupled, single-mode ($250 \: \mathrm{nm}$ top width) TFLN racetrack resonator. Forward- and backward-scanned resonance data is overlaid to show no significant photorefractive distortion. A Lorentzian fit gives an intrinsic quality factor of $Q_i \approx 5.1 \times 10^5$ (see Supplementary Section~\ref{si:ring} for details), corresponding to a propagation loss of $\alpha = 2.9 \pm 0.2$ dB/cm. (b) Finite-difference time-domain (FDTD) simulation of a $1\times2$ multimode-interference (MMI) splitter with an optimized multimode length $L_{\mathrm{MMI}} = 36 \: \mathrm{\mu m}$ (top) and directional coupler (bottom). (c) Fabricated MMIs are cascaded in a cutback configuration to determine the excess insertion loss per splitter, yielding an estimated value of 0.5 dB per MMI for $L_{\mathrm{MMI}} = 34 \: \mathrm{\mu m}$. Inset: measured grating-coupler efficiency as a function of grating duty cycle for an empirically optimized pitch of $274~\mathrm{nm}$. A minimum fiber-to-chip coupling loss of 4.8 dB is achieved. (d) Inset: as-fabricated directional-coupler power-splitting ratio as a function of the waveguide gap, showing large performance swings for nanometer-scale variations. Main plot: in situ tuning of a directional coupler using an integrated microheater.}
\label{fig2}
\end{figure}

Racetrack resonators are patterned in the TFLN layer to measure propagation loss. The laser (Daylight Solutions Stretto 493 nm ECDL) wavelength is tuned by applying a voltage to an internal piezo-electric controller. An imbalanced fiber MZI is constructed out of a pair of 50:50 fiber beamsplitters and a 1 meter path-length imbalance to record synchronous calibration fringes. The FSR of the MZI is itself calibrated against a Fabry--P\'erot interferometer (Thorlabs SA210-3B) with a known 10 GHz FSR, yielding an MZI FSR of 188 MHz. A variable optical attenuator (Thorlabs V520PA) is used to control the input optical power and thus mitigate the photorefractive effect. In this experiment, polarization-maintaining lensed fibers are used to couple to the chip [Fig.~\ref{fig2}(a)]. Single-mode (250 nm top width) racetrack resonators exhibit intrinsic quality factors of $Q_i \approx 5.1 \times 10^5$ and a corresponding propagation loss of $2.9 \pm 0.2$ dB/cm (Fig.~\ref{fig2}(a), bottom). For details on the quality factor measurement and loss extraction, see Supplementary Section~\ref{si:ring}.

Following validation of the passive waveguide, we develop integrated beamsplitters. In this work, we design two types of beamsplitters, $1 \times 2$ multimode-interference (MMI) splitters and evanescent directional couplers. FDTD simulation (Tidy3D) of a $1 \times 2$ MMI [Fig.~\ref{fig2}(b), top] reveals optimum multimode region dimensions of $36 \times 4 \: \mathrm{\mu m}$. Cutback measurements with cascaded components are performed to determine the insertion loss of each splitter and empirically optimize device parameters. Specifically, we  use a series of cascaded $1 \times 2$ MMI pairs, up to 60 MMIs (30 pairs). The optimum on-chip device has a multimode interaction length $L_{\mathrm{MMI}} = 34 \: \mathrm{\mu m}$ and excess insertion loss of 0.5 dB [Fig.~\ref{fig2}(c)]. Grating couplers, which are used for coupling light on and off the chip in the MMI study, reach a peak efficiency of $-4.8$ dB/grating for an optimized pitch of 274 nm and duty cycle of 55\% [Fig.~\ref{fig2}(c), inset]. However, we note the efficiency of grating couplers on thin TFLN stacks as used in this work is subject to wide performance swings owing to thickness non-uniformity (see Fig.~\ref{fig:S1} and Supplementary Section~\ref{si:nonuniform}).

FDTD simulations of an evanescent directional coupler show an optimum coupling gap of 250 nm and a coupling length of $43 \: \mathrm{\mu m}$ [Fig.~\ref{fig2}(b), bottom]. We characterize the fabricated couplers in a cascaded splitter tree, in which the through and cross ports are tapped out individually after successively more splitters, up to $N = 11$, to extract the per-splitter through and cross transmission. The splitting ratio is sensitive to nanometer-scale variations in the coupling gap [Fig.~\ref{fig2}(d), inset], so fabrication-induced deviations from the target ratio must be corrected after fabrication. We therefore tune the couplers in situ with integrated thermo-optic microheaters, which bring the splitting ratio back to its target. The heater consists of a thin, high-resistance NiCr layer ($R=1.4~\mathrm{k\Omega}$) deposited above the coupled-waveguide region. Joule heating differentially tunes the effective indices of the symmetric and antisymmetric modes, and thus the net power transfer over the coupling length. As shown in Fig.~\ref{fig2}(d), an initially imbalanced directional coupler is tuned through the 50:50 condition with modest electrical power of 3 mW.

\section{Electro-optic modulation}
\label{sec:modulators}

To fully leverage the properties of TFLN, we require high-bandwidth active components in addition to passive devices. We first characterize traveling-wave phase modulators, then apply the same electrode design to push--pull amplitude modulators.

\begin{figure}[htbp]
\centering\includegraphics[width=5.25in]{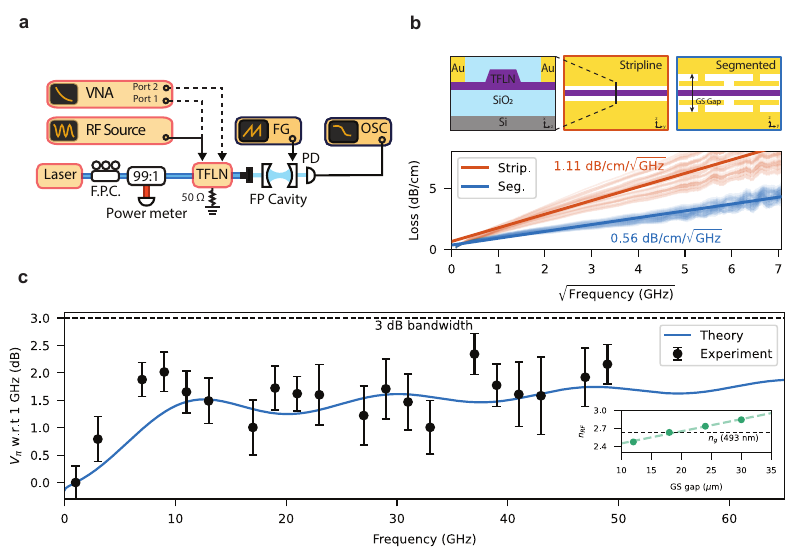}
\caption{\textbf{High-frequency electro-optic modulation at 493 nm.} (a) Experimental setup used to characterize the frequency-dependent response of the phase modulators. Light from a 493 nm laser is coupled to the TFLN chip through a fiber polarization controller (F.P.C.) and grating couplers. The modulated output is analyzed using a scanning Fabry--P\'erot (FP) cavity and photodetector (PD), while the tapped optical power is monitored with a power meter. FG, function generator; OSC, oscilloscope. For electrical S-parameter extraction, a VNA is connected (dashed line) to the modulator. (b) Cross section of the TFLN phase modulator and schematics of the stripline and segmented traveling-wave electrode geometries. Wafer-scale measurements of RF propagation loss for 29 stripline and 29 segmented-electrode devices are shown below. Solid lines denote the averages, demonstrating an approximately twofold reduction in frequency-dependent RF loss for the segmented electrodes. (c) Measured half-wave voltage relative to its value at 1 GHz for a phase modulator, together with the response predicted from electrical measurements. The dashed line marks a 3 dB increase in $V_{\pi}$ (a factor of $\sqrt{2}$). No 3 dB roll-off is observed within the measurement range, demonstrating an electro-optic bandwidth exceeding 50 GHz, limited by the available RF source. Inset: the velocity matching condition is achieved by tuning the gap between the main body of the ground and signal electrodes.}
\label{fig3}
\end{figure}

\subsection{Phase modulation}
To perform high-frequency characterization, continuous wave light from the 493 nm diode laser passes through a polarization controller and is coupled through a grating coupler onto a TFLN chip containing 3 mm long electro-optic modulators [Fig.~\ref{fig3}(a)]. The modulator is driven by an RF source (Keysight E8257D) and terminated with an off-chip $50~\Omega$ load to suppress microwave reflections. The modulated output is collected and analyzed using a scanning Fabry--P\'erot interferometer (Thorlabs SA210-3B). For electrical measurements, a 50 GHz VNA is connected to either end of the modulator electrodes to record the device's S-parameters.

To maximize the bandwidth of the TFLN EOM, we employ a segmented electrode design [Fig.~\ref{fig3}(b)], which has been shown to greatly reduce RF loss by mitigating Ohmic losses due to current crowding \cite{kharel2021breaking}. A wafer-scale comparison of 29 segmented and 29 stripline electrodes ($N=58$ devices in total) confirms this benefit, with the segmented design halving the average RF propagation loss from $1.11$ to $0.56$~dB\,cm$^{-1}$\,GHz$^{-1/2}$ [Fig.~\ref{fig3}(b)]. Residual variance in RF losses within segmented or stripline designs is a result of a parameter sweep of electrode geometry. 

In addition to reducing the RF losses, the segmented electrode design also provides a larger parameter space to achieve velocity matching. Efficient traveling-wave modulation requires both energy and momentum conservation, meaning the generated first-order sideband is displaced from the carrier by $\Delta \omega =\Omega_{RF}$ and $\Delta \beta=\beta_{RF}$, where $\Omega_{RF}$ and $\beta_{RF}$ are the microwave angular frequency and propagation constant, respectively. For modulation frequencies small relative to the optical carrier frequency, this condition reduces to the requirement that the optical group velocity match the microwave phase velocity, $v_{g,\mathrm{opt}} = v_{p,\mathrm{RF}}$, or equivalently that the optical group index $n_g$ match the effective microwave index $n_{\mathrm{RF}}$. Here, we tune the microwave effective index by varying the gap between the main ground and signal electrodes (GS gap, shown in Fig.~\ref{fig3}(b)), and the width of the signal electrode, which brings $n_{\mathrm{RF}}$ continuously through the optical group index, $n_g\approx2.63$ [Fig.~\ref{fig3}(c), inset]. Notably, these design variables achieve index control without substantially affecting the half-wave voltage, as the electrode spacing immediately around the waveguide remains constant (details on the segmented electrode design in Supplementary Section~\ref{si:electrodes}).

To overcome detector bandwidth limitations, we characterize the frequency-dependent electro-optic response of the segmented phase modulators through measurements of the generated optical sidebands. Under sinusoidal phase modulation, the output optical field can be expanded using the Jacobi--Anger identity:
\begin{equation}
    e^{i\omega_0t}e^{i\beta\sin(\Omega t)}
    =
    e^{i\omega_0t}
    \sum_{n=-\infty}^{\infty}
    J_n(\beta)e^{in\Omega t},
\end{equation}
where $\omega_0$ is the optical carrier frequency, $\Omega$ is the modulation frequency, and $J_n(\beta)$ is the Bessel function of the first kind. The modulation depth is
\begin{equation}
    \beta=\pi\frac{V_{\mathrm{RF}}}{V_{\pi}},
\end{equation}
where $V_{\mathrm{RF}}$ is the amplitude of the applied RF voltage and $V_{\pi}$ is the half-wave voltage of the phase modulator. The optical field amplitude of the $n$th sideband is proportional to $J_n(\beta)$, and its measured optical power is therefore proportional to $J_n^2(\beta)$.

At each modulation frequency, we sweep the applied RF power while simultaneously recording the output spectrum with the scanning Fabry--P\'erot interferometer. The measured carrier and sideband powers are fit to $J_n^2(cV_{\mathrm{RF}})$, where the fit parameter $c$ is inversely proportional to $V_{\pi}$. Normalizing $c$ to its value at 1 GHz yields the relative electro-optic response as a function of frequency. The measured response exhibits no 3 dB roll-off within the available measurement range, demonstrating an electro-optic bandwidth exceeding 50 GHz [Fig.~\ref{fig3}(c)]. For comparison, we calculate the expected response from the measured S-parameters and the simulated optical group index (Supplementary Section~\ref{si:electrodes}).

\begin{figure}[htbp]
\centering\includegraphics[width = 5.25in]{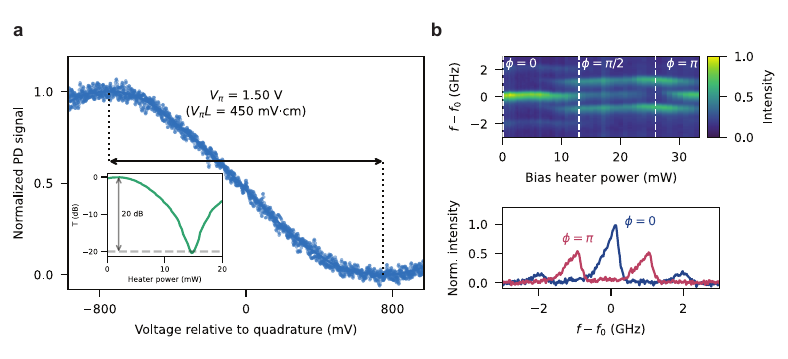}
\caption{\textbf{Amplitude modulation at 493 nm.} (a) Low-frequency (1 MHz) transfer function of a 3 mm long modulator, giving $V_\pi L = 450$ mV$\cdot$cm. Inset: the extinction ratio of an MMI-based, 3 mm long modulator is measured by sweeping the power delivered to an integrated microheater above one waveguide arm, reaching an extinction ratio of 20.4 dB. (b) Top: colormap of generated sidebands as a function of bias phase for a 1 GHz sinusoidal drive. Bottom: lineouts of the above plot. A balanced (null-biased) amplitude modulator preserves even (odd) order sidebands in the output.}
\label{fig4}
\end{figure}

\subsection{Amplitude modulation}
Push-pull, MMI-based amplitude modulators are characterized using a similar experimental setup, with an additional DC source used to control the thermo-optic bias through an integrated microheater. Driving a 3 mm long modulator at 1 MHz, we measure a half-wave voltage of $V_\pi = 1.50$ V, corresponding to a voltage-length product of $V_\pi L = 450$ mV$\cdot$cm for an electrode gap of 3~$\mu\mathrm{m}$ [Fig.~\ref{fig4}(a)]. By sweeping the heater bias, we measure the transfer function of the modulator and obtain a maximum extinction ratio of 20.4 dB [Fig.~\ref{fig4}(a), inset]. We attribute the residual transmission at the null point primarily to mode mixing and expect that narrower single-mode tapers preceding the active modulator section could further improve the extinction ratio.

The output spectrum of the push-pull Mach--Zehnder amplitude modulator can be understood as the coherent recombination of fields that undergo equal and opposite phase modulation in its two arms. Neglecting beam-splitter prefactors, the output field is
\begin{equation}
    E_{\mathrm{out}}(t)
    \propto
    e^{i\omega_0t}
    \sum_{n=-\infty}^{\infty}
    \left[
        J_n(\beta)
        +
        e^{i\phi}J_n(-\beta)
    \right]
    e^{in\Omega t},
\end{equation}
where $\beta$ is the phase-modulation depth in each arm and $\phi$ is the bias phase. Using the Bessel-function identity $J_n(-\beta)=(-1)^nJ_n(\beta)$, the amplitude of the $n$th sideband becomes
\begin{equation}
    E_n
    \propto
    J_n(\beta)
    \left[
        1+(-1)^n e^{i\phi}
    \right].
\end{equation}
The interference of each sideband at the output therefore depends on both the bias phase and the parity of its order. Figure~\ref{fig4}(b) shows the measured output spectrum under a 1 GHz drive as $\phi$ is swept using the integrated microheater. Selected spectra at $\phi=0$, corresponding to maximum transmission, and $\phi=\pi$, corresponding to the transmission null, are shown below. At maximum transmission, the odd-order sidebands are suppressed and only even orders remain. At the null, the carrier and other even-order components are suppressed, leaving only the odd-order sidebands. The resulting bichromatic fields are especially useful for Raman transitions and for simultaneous excitation of red and blue motional sidebands in trapped ion systems \cite{ransford2025helios}. The integrated modulator therefore provides a compact, single-pass approach to generating gigahertz-separated optical tones for atomic control. The apparent asymmetry of the laser lineshape persists when the TFLN PIC is bypassed and is a result of imperfect Fabry--P\'erot alignment, not the electro-optic device.

\section{Photorefractive response}
\label{sec:photorefraction}

\begin{figure}[htbp]
\centering\includegraphics[width=5.25in]{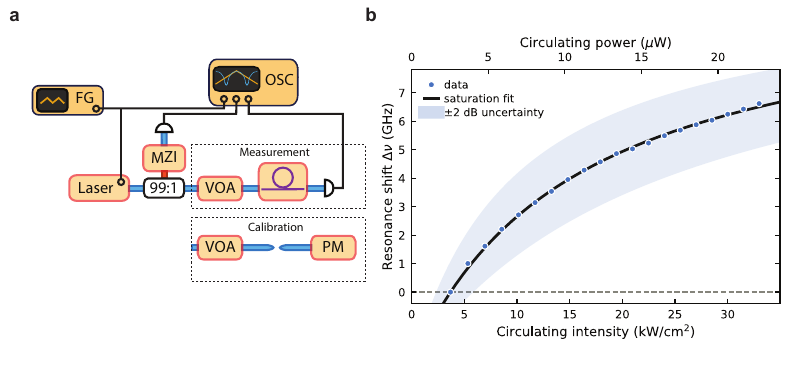}
\caption{\textbf{Photorefractive response at 493 nm.} (a) Experimental setup to probe the photorefractive response of the TFLN platform. Resonance frequencies are monitored under varying levels of incident optical power, while facet losses are estimated from a fiber-to-fiber reference path. (b) Measured resonance shift $\Delta\nu$ versus circulating intensity (top axis: circulating power), with saturation fit. The shaded band represents $\pm 2$ dB uncertainty in the intensity calibration, dominated by the facet-loss estimate.}
\label{fig5}
\end{figure}

An important limitation of electro-optic materials, particularly at short wavelengths, is their susceptibility to the photorefractive effect. Photorefraction occurs in multiple steps: first, carriers are optically promoted into conducting states from defect levels within the bandgap. The free charges then migrate by drift, diffusion, or the bulk photovoltaic effect, in which electrons preferentially move along the crystal $z$-axis owing to the anisotropic potential \cite{boyd2019nonlinear}. The resulting charge separation produces an electric field, which modifies the refractive index through the electro-optic effect. In TFLN, photorefraction is known to produce an index decrease (resonance blueshift) with increasing optical intensity, in contrast to the thermo-optic and Kerr effects \cite{boyd2019nonlinear, kukhtarev1978holographic}. Ring resonators allow us to observe photorefraction directly by measuring resonance shifts. As shown in Fig.~\ref{fig5}(a), we use a variable optical attenuator to tune the circulating intensity and monitor the resulting shift in equilibrium resonance frequency on an oscilloscope [Fig.~\ref{fig5}(b)]. The measurement is performed on the same single-mode resonance characterized in Fig.~\ref{fig2}(a); its linewidth, coupling condition, and FSR give a reliable estimate of the resonant enhancement factor. For details on the calculation, see Supplementary Section~\ref{si:ring}.

With this procedure we observe a photorefractive response together with the onset of saturation. The response is well described by the steady-state form $\Delta\nu = \Delta\nu_\mathrm{sat}\,I/(I+I_\mathrm{sat})$, where $\Delta\nu$ is referenced to the lowest measured intensity so that the fit includes a constant offset. A fit gives $\Delta\nu_\mathrm{sat}=13.4\pm0.1$~GHz and $I_\mathrm{sat}=15.7\pm0.6$~kW\,cm$^{-2}$ ($R^2=0.999$), corresponding to an initial slope $\Delta\nu_\mathrm{sat}/I_\mathrm{sat}=0.85$~GHz\,(kW\,cm$^{-2}$)$^{-1}$. Saturating behavior is consistent with first-order theories of photorefraction, in which photoconductivity increasingly counteracts the photovoltaic current at higher intensities. The index change and the space-charge field that produces it
can be written as
\begin{equation}
  \Delta n \;=\; -\tfrac{1}{2}\, n_e^{3} r_{33}\, E_{\mathrm{sc}},
  \qquad
  E_{\mathrm{sc}} \;=\; \frac{\kappa I}{\sigma_d + \beta I}
  \;\xrightarrow{\;\beta I \,\gg\, \sigma_d\;}\; \frac{\kappa}{\beta},
\end{equation}
where $\kappa I$ is the photovoltaic current density, $\sigma_d$ the dark
conductivity, and $\beta I$ the photoconductivity~\cite{glass1995high,katz2004vapor}. Where photoconductivity dominates, the steady-state space-charge field becomes
independent of intensity. However, the power regime over which this study was conducted likely does not capture all the complexities of the photorefractive phenomenon. Furthermore, the effect is known to depend on specific fabrication processes, such as the presence of a top cladding layer \cite{xu2021mitigating}, which is a desirable part of our modulator process. Further work is warranted to study the photorefractive effect in TFLN over a larger range of input optical powers and wavelengths, and it remains a critical factor in choosing between different electro-optic platforms.

\section{Discussion and conclusion}

We have demonstrated a set of passive and active photonic components operating at 493 nm on thin-film lithium niobate, fabricated at the wafer scale. Single-mode waveguides exhibit a propagation loss of $2.9 \pm 0.2$ dB/cm; MMI beamsplitters an excess loss of 0.5 dB; and grating couplers a peak efficiency of $-4.8$ dB, while integrated heaters tune directional couplers through the 50:50 point. Modulators based on segmented traveling-wave electrodes combine a voltage-length product of $V_\pi L = 450$ mV$\cdot$cm, an electro-optic bandwidth exceeding 50 GHz (limited by the available RF source), and an extinction ratio of 20.4 dB. Together, these results show that the gigahertz-scale electro-optic control needed to address atomic hyperfine structure can be realized on an integrated platform at a blue ion wavelength.

This work is ultimately motivated by efforts towards realizing scalable trapped-ion systems. Roadmaps toward fault-tolerant trapped-ion and neutral-atom processors call for thousands to millions of physical qubits, each requiring several independently controlled optical channels, a scale at which free-space delivery and bulk modulators become impractical and integrated photonics is widely expected to be necessary \cite{park2024technologies}. Integrated devices have started to deliver light to trapped ions and enable on-chip cooling and photon collection \cite{niffenegger2020integrated, mehta2020integrated, corsetti2026integrated, knollmann2025collection}, and integrated modulators have driven high-fidelity qubit operations, although with lower stability than the bulk acousto-optic modulators they replace \cite{hogle2023high}. While ambitious efforts aim to co-integrate complete optical systems with ion traps, we anticipate that the near-term priority is the stability and reliability of simpler building blocks such as the modulators presented here. Their requirements differ from those of telecommunications, where low drive voltage and energy per bit dominate \cite{wang2018integrated}. For quantum control, extinction ratio and long-term stability, in particular at the different bias points used in practice, matter most, because they set gate errors, crosstalk, and calibration overhead. Similar considerations apply to other technologies that rely on fast control of visible light, from neutral-atom arrays and optical clocks to underwater communication \cite{park2024technologies}.

At the device level, higher extinction should be attainable with single-mode tapers that suppress the mode mixing limiting our devices, or with cascaded modulator stages. Long-term stability remains an open question. Photorefraction and DC bias drift are more pronounced at short wavelengths and must be characterized across operating bias points, and their mitigation through cladding engineering and doping warrants systematic study \cite{celik2024roles, xu2021mitigating, suntsov2026photorefractive}.


Materials and fabrication pose further challenges. The TFLN market is driven by telecommunication devices, and commercial wafers are typically derived from congruent material, which is lithium deficient. The resulting antisite defects act as electron traps that contribute to photorefraction and light-induced absorption \cite{schirmer2009electron}. Near-stoichiometric crystals contain fewer such defects \cite{gopalan1998role} and, when lightly doped with MgO, strongly suppress photorefractive damage \cite{furukawa2000photorefraction}, so dedicated materials development toward more stoichiometric, damage-resistant thin films would directly benefit visible operation. Thickness uniformity is equally important. A fixed absolute variation in film or slab thickness represents a larger fractional variation in the thinner layers required here, and even in thicker films, wafer-scale variations in thickness and etch depth can shift nominally identical devices far from their design targets \cite{xin2025wavelength}. The strong sensitivity of grating couplers to film thickness variation (Fig.~\ref{fig:S1}) illustrates the same effect at 493 nm, so thickness-aware design, in which wafer metrology is fed back into the device layout \cite{xin2025wavelength, chen2024adapted}, together with wafer-scale trimming \cite{kopparapu2026deterministic}, will be important for yield. Lithography is another constraint. Our devices use sub-200-nm features, such as grating teeth and gaps, which are readily achievable by electron-beam lithography; however, these dimensions are near the resolution limit of the ultraviolet lithography used in most photonic foundries, complicating a direct transfer to foundry processes.

Packaging is a further hurdle. The small mode sizes of visible waveguides make fiber coupling highly sensitive to misalignment, and grating couplers currently dominate our optical loss budget. Scalable interconnect technologies such as photonic wire bonding have been developed primarily for near-infrared and telecommunication wavelengths \cite{blaicher2020hybrid}, and their performance at blue wavelengths remains largely unexplored. Finally, heterogeneous integration of several platforms will be beneficial, combining silicon nitride resonators and extended cavities to narrow and stabilize visible lasers \cite{corato2023widely, song2025integrated, castro2025integrated}, silicon nitride and alumina for low-loss routing and co-integration with ion traps \cite{west2019low, corato2024absorption, niffenegger2020integrated, mehta2020integrated}, and electro-optic thin films, including lithium tantalate at shorter wavelengths \cite{lin2026thin, wang2024lithium}, for fast modulation. Progress in device stability and extinction, wafer quality and uniformity, foundry-compatible fabrication, and visible-wavelength packaging, combined with such hybrid integration, outlines a practical path from the component-level results presented here toward scalable optical control of atomic systems.

\section*{Funding}
DRS Daylight Solutions and Monarch Quantum, Inc. (NA, XL, AS), Capes-Fulbright and Behring foundation fellowships (LM), NSF ERC Center for Quantum Networks EEC-1941583 (HW, SD, XZ), Rubicon postdoctoral fellowship from the Dutch Research Council (019.231EN.011), Gordon and Betty Moore Foundation (AC), NSF GRFP No. DGE1745303 (DR, DA), NSF OMA-2137723 (AZ)

\section*{Disclosures}
M.L., D.R., and H.W. are involved in commercializing TFLN technologies at HyperLight Corporation. M.L. and K.P. are involved in commercializing TFLT technologies at Lumina Corporation. A.S-A and M.P. are involved in building quantum light engines at Monarch Quantum. The remaining authors declare no conflicts of interest.

\section*{Data availability}
Data underlying the results presented in this paper are not publicly available at this time but may be obtained from the authors upon reasonable request.


\clearpage
\begin{center}
{\Large\bfseries Supplementary Information}\\[4pt]
{\large Thin-film lithium niobate integrated photonics at 493\,nm}
\end{center}

\setcounter{section}{0}
\setcounter{figure}{0}
\setcounter{equation}{0}
\setcounter{table}{0}
\renewcommand{\thesection}{S\arabic{section}}
\renewcommand{\thefigure}{S\arabic{figure}}
\renewcommand{\theequation}{S\arabic{equation}}
\renewcommand{\thetable}{S\arabic{table}}
\renewcommand{\theHsection}{S\arabic{section}}
\renewcommand{\theHfigure}{S\arabic{figure}}
\renewcommand{\theHequation}{S\arabic{equation}}
\renewcommand{\theHtable}{S\arabic{table}}
\section{Effect of substrate nonuniformity}\label{si:nonuniform}
A significant difficulty in wafer-scale fabrication for short-wavelength photonics is the nonuniformity of the substrate. Commercial TFLN wafers often suffer from 10--20~nm of total thickness variation (TTV) regardless of the nominal film thickness, so the thin stacks used in visible photonics suffer a larger fractional variation. This nonuniformity significantly degrades devices that rely on phase matching, such as the grating couplers used in this work. Nominally identical grating couplers ($N = 40$) sampled over a 16~cm$^2$ wafer area exhibit a wide spread in efficiency [Fig.~\ref{fig:S1}(a)], and none of them on the measured wafer reached our previous optimum of 4.8~dB, emphasizing the strong dependence on local film thickness. We verify this sensitivity using a 2D FDTD model in Fig.~\ref{fig:S1}(b). The simulated efficiency falls off rapidly around the optimum film thickness, reaching dB scale degradations for nanometers of film thickness away from the target design.

\section{Traveling-wave electrodes}\label{si:electrodes}
\begin{figure}[htbp]
\centering
\includegraphics[width=5.25in]{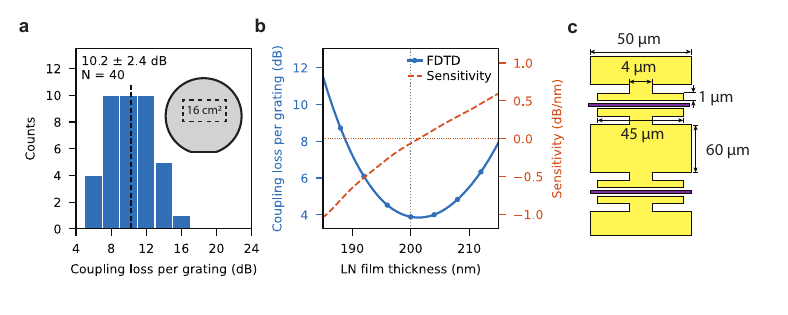}
\caption{\textbf{Grating-coupler variation across the wafer and electrode geometry.}
(a) Histogram of the measured coupling loss per grating for $N = 40$ nominally
identical grating couplers sampled over an approximately $57 \times 28$~mm
($\approx 16$~cm$^2$) region of the wafer (inset). The dashed line
marks the mean, $10.2 \pm 2.4$~dB (mean $\pm$ one standard deviation).
(b) Simulated (2D FDTD) coupling loss per grating as a function of LN film
thickness for a 125~nm etch depth (left axis), and its thickness sensitivity (right axis). Away from
the optimum, the sensitivity reaches $\approx 1$~dB/nm, so film-thickness
variations of a few nanometers shift the coupling loss by several dB.
(c) Top-view schematic (not to scale) of the segmented traveling-wave electrode (yellow)
alongside the optical waveguides (purple), with the key dimensions labeled.}
\label{fig:S1}
\end{figure}

The electrodes are designed with COMSOL and Ansys HFSS. COMSOL provides simultaneous RF and optical mode solving to estimate the microwave--optical overlap and the resulting half-wave voltages, while HFSS is used to predict the RF indices of the segmented electrode designs \cite{kharel2021breaking}. A typical segmented geometry is shown in Fig.~\ref{fig:S1}(c). Prior to optical testing, the fabricated traveling-wave electrodes are characterized with a vector network analyzer (VNA) to extract frequency-dependent S-parameters up to 50~GHz. We extract the microwave index from the slope of the accumulated phase of $S_{21}$ versus frequency, while its amplitude gives the microwave loss. Following electrical characterization, we select the best-performing devices for thorough optical testing. RF cable losses are calibrated out by measuring the cable loss directly and adjusting the delivered RF power accordingly during fitting.

These electrical measurements also predict the electro-optic response. The normalized frequency-dependent response of a traveling-wave electro-optic modulator can be written as
\begin{equation}
m(\Omega)
=
\left|
\frac{2Z_{\mathrm{in}}}{Z_{\mathrm{in}}+Z_G}
\,
\frac{
    (Z_L+Z_0)F_{+}
    +
    (Z_L-Z_0)F_{-}
}{
    (Z_L+Z_0)e^{\gamma_m L}
    +
    (Z_L-Z_0)e^{-\gamma_m L}
}
\right|,
\label{eq:eo_response}
\end{equation}
where $\Omega$ is the microwave angular frequency, $Z_G$ is the source impedance, $Z_0$ is the characteristic impedance of the transmission line, $Z_L$ is the load impedance, $\gamma_m$ is the complex propagation constant of the microwave mode, and $L$ is the interaction length \cite{ghione2009semiconductor}. The factors
\begin{equation}
F_{\pm}
=
\frac{
    1-
    \exp\left(
        \pm\gamma_mL
        -
        i\frac{\Omega n_gL}{c}
    \right)
}{
    \pm\gamma_mL
    -
    i\frac{\Omega n_gL}{c}
}
\label{eq:velocity_mismatch}
\end{equation}
describe the accumulated microwave--optical phase mismatch along the modulator, where $n_g$ is the optical group index and $c$ is the speed of light in vacuum. Microwave attenuation is contained in the real part of $\gamma_m$, while the difference between the microwave phase velocity and the optical group velocity determines the relative phase accumulated by the two fields. The effective input impedance of the terminated transmission line is
\begin{equation}
Z_{\mathrm{in}}
=
Z_0
\frac{
    Z_L+Z_0\tanh(\gamma_mL)
}{
    Z_0+Z_L\tanh(\gamma_mL)
}.
\label{eq:input_impedance}
\end{equation}
We extract $Z_0$ and $\gamma_m$ from the VNA measurements and obtain $n_g$ from optical-mode simulations. Substituting these quantities into Eqs.~(\ref{eq:eo_response})--(\ref{eq:input_impedance}) yields the theoretical electro-optic response shown in Fig.~\ref{fig3}(c) of the main text.

\section{Ring resonator measurements}\label{si:ring}
Owing to the small tuning range of the source laser, we fabricate TFLN racetrack resonators with small (30~GHz) free spectral ranges (FSRs) to increase the resonance density. The straight segments of the racetracks are oriented along the crystal y-axis, which influences their photorefractive response. Resonances are fit to a Lorentzian model with a linear background,
\begin{equation}
  T(\delta) = c_0 + c_1(\delta-\delta_0)
            - D\,\frac{(\gamma/2)^2}{(\delta-\delta_0)^2+(\gamma/2)^2},
\end{equation}
where $\delta$ is the detuning, calibrated against a fiber MZI. The fitted linewidth gives the loaded quality factor directly, $Q_L = \nu_0/\gamma$, with $\nu_0 = c/\lambda$ and $\lambda = 493$~nm. The intrinsic quality factor then follows from the contrast $T_\mathrm{min}$ and $1/Q_L = 1/Q_i + 1/Q_e$,
\begin{equation}
  Q_i = \frac{2Q_L}{1+\sqrt{T_\mathrm{min}}},
\end{equation}
taking the undercoupled branch ($Q_i < Q_e$), from which the propagation loss is $\alpha = 2\pi n_g/(Q_i\lambda)$. Finally, the intracavity intensity is $I = B\,P_\mathrm{bus}/A_\mathrm{eff}$, where $A_\mathrm{eff}$ is the simulated effective mode area and $B$ is the resonant enhancement factor \cite{haus1984waves},
\begin{equation}
  B = \frac{P_\mathrm{circ}}{P_\mathrm{bus}} = \frac{2}{\pi}\,F\,\frac{Q_L}{Q_e},
  \qquad F = \frac{Q_L\,\mathrm{FSR}}{\nu_0}.
  \label{eq:buildup}
\end{equation}

\end{document}